*Article*

# Disturbance Effects on Financial Timberland Returns in Austria

**Petri P. Kärenlampi [1\*]**

1   Lehtoi Research, Finland
\*   Correspondence: petri.karenlampi@professori.fi

**Abstract:** Probability theory is applied for the effect of severe disturbances on the return rate on capital within multiannual stands growing crops. Two management regimes are discussed, rotations of even-aged plants on the one hand, and uneven-aged semi-stationary state on the other. The effect of any disturbance appears two-fold, contributing to both earnings and capitalization. Results are illustrated using data from a recently published study, regarding spruce (*Picea abies*) forests in Austria. The economic results differ from those of the paper where the data is presented, here indicating continuous-cover forestry is financially inferior to rotation forestry. Any severe disturbance may induce a regime shift from continuous-cover to even-aged forestry. If such a regime shift is not accepted, the disturbance losses reduce profits but do not affect capitalization, making continuous-cover forestry financially more sensitive to disturbances. Revenue from carbon rent favors the management regime with higher carbon stock. The methods introduced in this paper can be applied to any dataset, regardless of location and tree species.

**Keywords:** gross profit; capitalization; capital return rate; expected value; probability density.

**Introduction**

Recently, a combination of drought and bark beetles has induced large salvage cuttings of forests in Central Europe [1, 2, 3]. Extended drought also has induced an increased risk of a wildfire [4, 5, 6]. Also combinations of windfall and bark beetles have been devastating [7, 8]. A European report however indicates insect damage and wildfires have diminished from year 2000 to 2015, whereas snow and wind damage have increased [9]. Such stand-replacing disturbances obviously are worth considering when comparing different kinds of management systems.

Severe disturbances on forest stands are often induced by an interplay of factors like drought, fire, wind, snow, landslides, floods, volcano eruptions, insects, pathogens, and sometimes mammals [10, 11, 12, 13, 14, 15, 16, 17]. Not all large-scale disturbances need to be stand-replacing [18, 19]. Wind damage rates may significantly differ between decades [20]. Some authors consider mortality events removing several percentages of trunk volume as catastrophic, appearing as an interplay with background mortality [21, 22]. Within the boreal region, the dominant stand-replacing disturbance is fire, with a natural return interval of decades or centuries [23, 24, 25, 26, 27, cf. 28]. The importance of windthrow appears to increase towards the temperate zone, often interacting with bark beetles [29, 30, 8, 31, 32]. Prolonged interval between stand-replacing disturbances appears to lead to gap regeneration [33, 18, 34, 35, 25, 19, 36, 37, 38]. Different mechanisms tend to control gap initiation and gap expansion [25].

The recovery rate of a system from any disturbance naturally depends on the kind of disturbance. Losses of growing stock in the context of forest stand thinning (a non-stand-replacing disturbance) are the fastest recovered in the case of young stands with relatively low density, also affected by precipitation and temperature [39].

Another question is financial performance in the occurrence of disturbances. Several investigations discuss mortality of forest trees in terms of stem counts [40]. Neglecting the size of dying trees however impairs the applicability of such studies, from the viewpoints of stand value and stand productive capacity, as mortality of individual trees due to within-species competition concentrates in small stems, as well as trees with slow growth rate [41, 42, 43, 19, 44, 45, 46, 47, 48, 49, 50, 51, 52, 15, 53]. However, tree species differ with respect to factors dominating mortality [54, 55, 56, 57, 53, 58]. With increasing tree size, physical mortality drivers appear to increase with the expense of biotic reasons [35, 59]. Dying trees may or may not remain standing, and their positions may be clustered [37, 60]. Aging of stands appears to increase sensitivity to disturbances [61, 62, 63, 64]. Young forests and small trees however appear more sensitive to heat and drought than old forests and large trees [65, 66, 67]. On the other hand, adverse weather conditions may have long-term effects, possibly interacting with pathogens, insects, and competition [68, 13, 48, 45, 69, 70, 71, 72, 73]. Disintegration of functional processes has been proposed as a causal explanation [74]. Hot draughts are more lethal than temperate ones [75]. Some observations indicate low temperatures induce mortality [67]. Relationships between forest stand observables and demographic processes change along with time, possibly due to changing climate [76, 11, 12, 30, 77, 69, 78, 79].

A severe, unintended disturbance may or may not require cultivation with new seedlings. If it does require cultivation in the case of continuous-cover forestry, at least a temporary regime shift into even-aged forestry takes place. In even-aged rotation forestry, regenerative disturbances do not induce any regime shift. However, the crop may be lost, completely or partially.

The time needed to recover from a regenerative disturbance can be addressed as follows. Within rotation forestry, the expected value of recovery time from a realized regenerating disturbance is

$$\langle RT \rangle = \int_0^\tau k(a) a \, da \qquad (1),$$

where $k(a)$ is the probability density of realizing regenerating disturbances within the rotation time $\tau$. In the possible case of evenly distributed regenerating disturbances, the expected value of the recovery time would approach half of the rotation time – other kinds of distributions also are feasible [61, 62, 2].

Figure 5a of Knoke et al. [80] indicates that after a regenerating disturbance, the recovery time into the semi-stationary state of continuous-cover forestry is 60 years at least, corresponding to one full rotation within the even-aged management.

Financial performance indeed is another subject. Fig. 5a of Knoke et al. [80] indicates that the financial performance of the continuous-cover forestry impairs as the semi-stationary state is approached. Correspondingly, the recovery time does not have any direct relation to financial performance, which is here approached in terms of return rate on capital.

The economic value of a forest is given by Knoke et al. [80] as

$$C(b) = \int_b^\infty \frac{dR}{dt} \exp(-rt) dt \quad (2),$$

where $\frac{dR}{dt}$ is the time rate of gross profits, $r$ is a discount rate, and $b$ is the present time. Such an estimate of the economic value strongly depends on the discount rate. Another complication is that the gross profit rate varies not only with time, but also with a variety of factors, including forest age. The latter complication can be overcome by integrating the gross profit rate over stand age distribution, to produce an expected value

$$\left\langle \frac{dR}{dt} \right\rangle(b) = \int_0^\tau p(a) \frac{dR}{dt} da \quad (3),$$

where $a$ is stand age, $p(a)$ is its probability density, and $\tau$ is rotation age.

At this stage the discount rate $r$ is unknown. However, in developed markets, the forest value C is known as a market value. Then, the expected value of the discount rate can be readily resolved from Eq. (2),

$$\langle r \rangle(b) = \frac{\left\langle \frac{dR}{dt} \right\rangle(b)}{\langle C \rangle(b)} \quad (4),$$

where it gains the interpretation of *the expected value of capital return rate* at time $b$. Such an expected capital return rate may apply to a single stand by the collapse of the probability density functions. On the other hand, being determined on the basis of the market values of forest estates, it can be taken as the *capita return rate requirement* on the market.

In the remaining part of this paper, we will develop produces for discussing the effects of stand-replacing disturbances on the expected capital return rates in terms of probability theory. The question is nontrivial since disturbances contribute to both gross profit rate (Eq. (3)) and to capitalization (Eq. (2)). The procedures to be developed are supposed to be universal – results however will be produced regarding two management systems of spruce forests in Austria, using data from the recent publication from Knoke et al. [80]. The two management regimes discussed here are
- even-aged forestry with artificial regeneration
- continuous-cover forestry with uneven-aged structure.

It is worth noting that the continuous-cover forestry discussed here differs from the system of prolonged rotations with partial artificial and partial natural regeneration discussed by Knoke et al. [80].

**Materials and methods**

Regenerative disturbances inducing the loss of crops have financial implications. The probability density of stand age appearing in Eq. (3) certainly is affected, and it affects not only the gross profit rate, but also the capitalization. The expected value of capitalization appearing in Eq. (4) can naturally be written

$$\langle C \rangle = \int_0^\tau p(a) C \, da \quad (5).$$

Then, provided *p(a)* is taken as the probability density of stand age in the absence of regenerating disturbances, the probability density in the presence of disturbances becomes

$$p'(a) = p(a), \text{ if } \frac{dM}{dt} = 0$$

$$p'(a) = \frac{\frac{dM}{dt} \exp(-a \frac{dM}{dt})}{1 - \exp(-\tau \frac{dM}{dt})}, \text{ if } \frac{dM}{dt} > 0 \qquad (6),$$

where $\frac{dM}{dt}$ is the rate of regenerating disturbance density, $\exp(-a \frac{dM}{dt})$ is the probability of stand survival to age *a*, and $\frac{\frac{dM}{dt}}{p(a)\left[1 - \exp(-\tau \frac{dM}{dt})\right]}$ is the probability density normalization factor. The normalization factor becomes like this provided the non-disturbed probability density of stand age $p(a)$ is taken as a constant, corresponding to evenly distributed stand ages. The effect of disturbances can be straightforwardly addressed by substituting Eq. (6) into Eqs. (3) and (5), and further to Eq. (4).

A question arises whether the above Equations are applicable to continuous-cover forestry. Eqs. (2), (3), (4), and (5) indeed are applicable, but the rotation time $\tau$ as a parameter deserves some discussion. In semi-stationary continuous-cover forestry, an operative rotation time corresponds harvesting rotation, rather than the life cycle of trees. Equations (1) and (6) discuss regenerating disturbances. In the case of continuous-cover forestry, any regenerating disturbance induces a temporary shift of management regime, at least if regeneration is aided in terms of cultivation.

It is, however, possible to apply the rate of regenerating disturbance density $\frac{dM}{dt}$ to continuous cover forestry, assuming that the disturbance is distributed and does not become compensated by cultivation. One can then assume that the disturbance loss rate $\frac{dM}{dt} C$ is deducted from the gross profit rate, resulting as disturbance-affected gross profit rate

$$\left(\frac{dR}{dt}\right)' = \frac{dR}{dt} - \frac{dM}{dt} C \qquad (7).$$

It is here assumed that provided $\frac{dR}{dt} \geq \frac{dM}{dt} C$, any disturbance only reduces the gross profit rate, and does not affect the capitalization within the continuous-cover regime.

Let us then turn to the recent publication of Knoke et al. [80] for the clarification of input variables characteristic for spruce forest in Austria. Firstly, Table 2 of [80] indicates a regeneration expense of 2000 Eur/ha. Fig. (5) of [80] indicates a bare land value of 10000 Eur/ha, and the value of trees of 18000 Eur/ha at the age of 60 years. Within this range, the Figure implies an expected value of capitalization of 21000 Eur/ha. Regarding the continuous-cover forestry, Fig. 5a indicates a semi-stationary state capitalization of 24000 Eur/ha. Table 2 indicates a gross profit rate of 154 Eur/(ha*a) at the semi-stationary state.

The publication of Knoke et al. [80] does not indicate the volumetric timber stocks, or stored amounts of $CO_2$. Given the 3000 Eur/ha difference in the expected value of the capitalization among the management regimes, one can approximate a timber stock difference in the order of 50 m³/ha.

Within the regime of even-aged rotation forestry, the expected value of the gross profit rate is naturally produced using Eq. (3), and that of capitalization according to Eq. (5).

**Results**

Figure 1 shows the expected value of capital return rate as a function of disturbance density rate $\frac{dM}{dt}$ for the two regimes. In the absence of disturbances the capital return rate of even-aged rotation forestry is two times that of the continuous-cover forestry. The continuous-cover forestry is much more sensitive to disturbances, as any expected annual damage loss directly reduces gross profit according to Eq. (7): within the range of disturbance densities appearing in Fig. 1, the disturbances reduce the expected gross profit up to 86%, and the capital return rate correspondingly. The effect of disturbances on the expected value of capital return rate within the even-aged rotation forestry is much gentler, diminishing only 20% in Fig. 1.

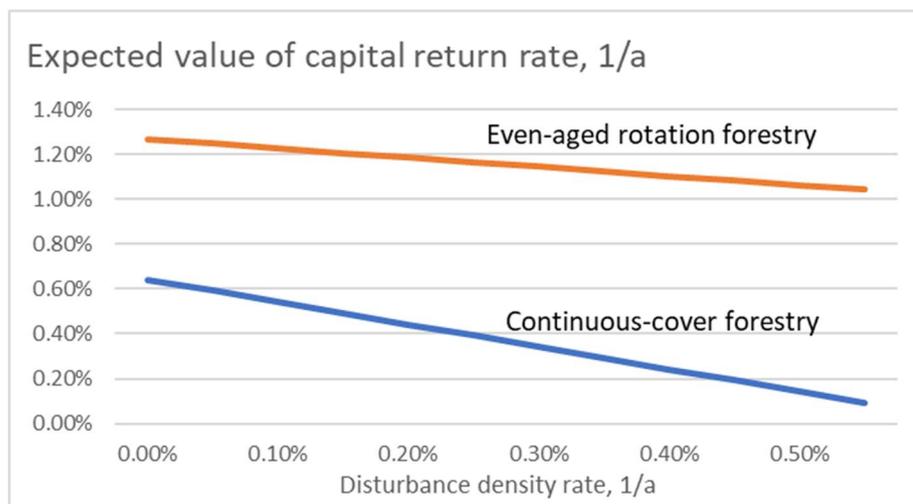

Figure 1. The effect of disturbances on the expected value of capital return rate for the two regimes.

**Discussion**

As the effect of disturbances on the finances of the continuous-cover forestry is rather simple according to Eq. (7), the even-aged rotation forestry case is a more delicate subject of discussion, containing a change in the gross profit rate, change in capitalization, change in survival probability of individual stands, as well as probability density of stand age.

Figure 2 shows that the capitalization decreases slightly, according to Eq. (5). The capital return rate reduces at most 18%, and the gross profit rate at most 20%. The difference between the change in the capital return rate and the gross profit rate is due to the change in capitalization, according to Eq. (4).

Interestingly, the biggest change due to disturbances is in the survival probability of stands until final harvesting, at most -28% (Fig. 2). Considering that positive revenue is gained only from final harvesting (cf. Fig. 5 of [80]) one might ask why the gross profit rate is not as much affected as the stand survival probability. The answer naturally lies in the probability density of stand age.

In Figure 3, the appearance density of mature stands is reduced by 16%, even if the survival probability of individual stands is reduced by 28%. The reason for this difference is that the probability density of newly established stands increases by 17% along with disturbances. Correspondingly, there is a greater number of stands regenerated annually. The probability density of mature stands can be verified as the product of that of newly regenerated stands, multiplied by the survival probability.

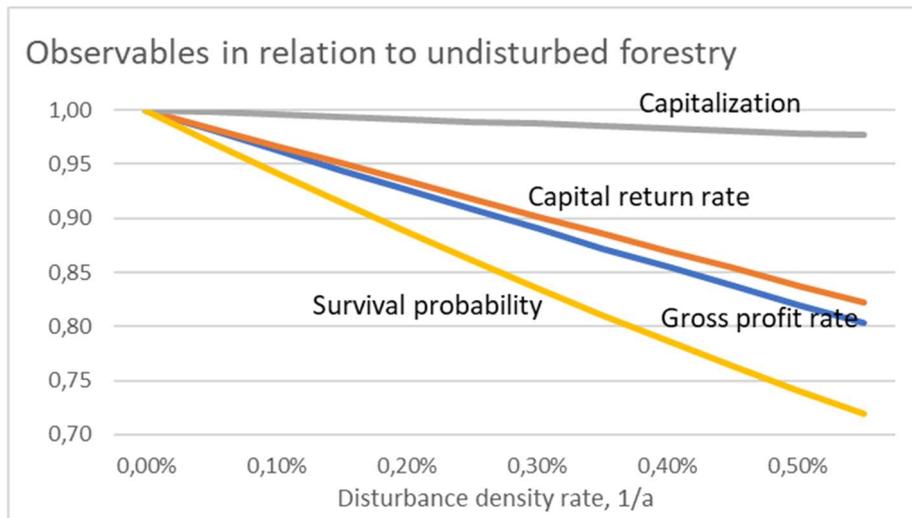

Figure 2. The effect of disturbances on capitalization, capital return rate, gross profit rate, and survival probability in even-aged rotation forestry.

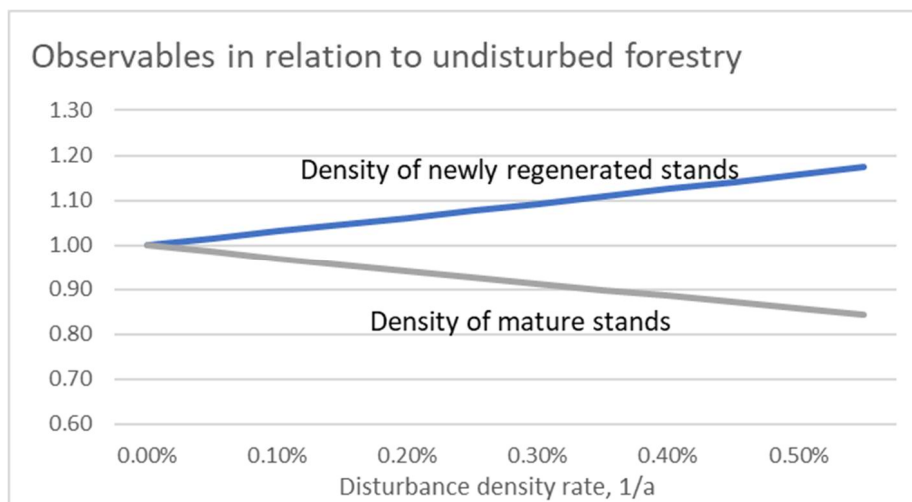

Figure 3. The effect of disturbances on the change in appearance density of newly regenerated stands and stands mature for harvesting.

An issue worth discussing is the applicability of the two management regimes in carbon sequestration and storage. Within the two implementations described by Knoke et al. [80], the semi-stationary state in the continuous-cover forestry has greater capitalization and greater timber stock, in relation to the expected values within rotation forestry. The greater timber (or carbon) stock may become compensated in terms of a carbon rent. An additional carbon stock of 50 tons/ha, compensated with a carbon rent of 2 Eur/(ton*a), would increase annual revenue within the CCF by 100 Eur/ha. Such additional revenue would improve the capital return rate within CCF, and Fig. 1 would become modified as shown in Fig. 4. It is found that the capital return rate within the CCF still is inferior to that of RF, and it still is more sensitive to disturbances. Obviously, an increment of the carbon stock could be aspired within the rotation forestry. Details of such an attempt however are beyond the scope of this study.

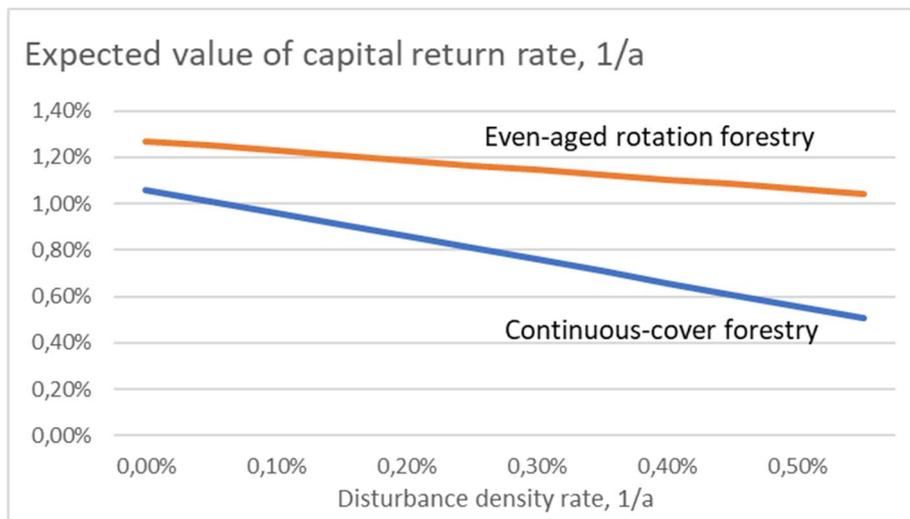

Figure 4. The effect of disturbances on the expected value of capital return rate for the two regimes, when the greater carbon stock of the CCF is compensated by a carbon rent of 100 Eur/(ha*a).

The present treatment is extremely simplified. First, one might ask whether it induces any bias that the non-disturbed probability density of stand age for undisturbed forest $p(a)$ was taken as a constant. Strictly, the constancy would refer to an assumption of a *normal forest* – structure [81]. Such an assumption however is not necessary; any single stand can be observed at a random time instant within any rotation, corresponding to a constant probability density of stand age [82, 83]. On the other hand, the procedures above allow any analyst to repeat the procedure with appropriate probability density of stand ages.

Secondly, one might ask what are the consequences of assuming a constant regenerating disturbance density rate $\frac{dM}{dt}$ with respect to stand age. Aging is known to make forests more vulnerable [61, 62, 63, 64]. However, the rotation ages applied here are probably not long enough for such effects to materialize. Again, any analyst is invited to insert desired age-dependent disturbance rates.

A third question is the definition of the "regenerating disturbance". Intimately related to this is any eventual revenue from salvage harvesting. In Eq. (6), a regenerating disturbance nullifies the local stand age and induces a regeneration expense. Any salvage revenue is omitted. In reality, there often is salvage revenue. On the other hand, there are disturbances not resulting in regeneration. The justification for simply omitting any salvage revenue is that such a procedure compensates the omitted loss due to non-regenerating disturbances. The same goes with continuous-cover forestry: omitted salvage revenues compensate any non-salvaged non-regenerating damage.

There is no guarantee that the data used in this study would be representative of spruce forests in Austria. In particular, it is questionable whether the cross profits, as well as capitalizations appearing in the semi-stationary state of continuous-cover forestry are sustainable [84, 85]. One must realize that the data used here describes one example case. The methods used above can be readily applied to any available set of cross profits and capitalizations, corresponding to different management regimes.

The outcome regarding financial resilience in the occurrence of disturbances is very different from that of Knoke et al. [80]. An obvious reason is that the present treatment is purely financial, whereas Knoke et al. [80] applied a mixture of economical and time-delay criteria. Another major difference is that we have here discussed continuous-cover forestry as one management regime, instead of prolonged rotations with partially artificial and partially natural regeneration.

**Conclusions**

Procedures were developed for the assessment of financial consequences of severe disturbances on multiannual stands growing crops, in terms of probability theory. Two management regimes were discussed, rotations of even-aged plants on the one hand, and uneven-aged semi-stationary state on the other. The effect of any disturbance appeared two-fold, contributing to both earnings and capitalization. Results were illustrated using data from a recently published study, regarding spruce (*Picea abies*) forests in Austria. The economic results differed from those of the paper where the data was presented, here indicating continuous-cover forestry is financially inferior to rotation forestry. Any severe disturbance may induce a regime shift from continuous-cover to even-aged forestry. If such a regime shift is not accepted, the disturbance losses reduce profits but do not affect capitalization, making continuous-cover forestry financially more sensitive to disturbances. Revenue in terms of a carbon rent favors the management regime with higher carbon stock. The methods introduced in this paper can be applied to any dataset, regardless of location and tree species.

**Funding:** This research was partially funded by Niemi Foundation.

**Data Availability Statement:** Datasets used have been introduced in earlier papers referenced above.

**Conflicts of Interest:** The author declares no conflict of interest.